\documentclass[aps,eqsecnum,twocolumn,prb]{revtex4}
\usepackage{graphics}

\begin{document}
\begin{center}

\textbf{\large Accurate electronic excitations for two alkali-halide systems obtained by density-functional theory and verified by multi-configuration self-consistent field calculations}

A.~Hellman\footnote{Corresponding author: Tel. +46 31 772 3377;
Fax: +46 31 772 8426; E-mail: \texttt{ahell@fy.chalmers.se}} and M.~Slabanja\\
Department of Applied Physics,
 Chalmers University of Technology and G\"oteborg University, 
 SE-412 96 G\"oteborg, Sweden.
\end{center}

\vspace{\parskip}\textbf{Abstract}

Use of density-functional theory in a $\Delta$self-consistent field
framework result in both the ground- and two lowest electronicly excited
states of the NaCl and LiCl. The accuracy of this method is confirmed
using a multi-configuration self-consistent field method to obtain the
same states. The overall good agreement between the calculated ground
and excited potential-energy surfaces speaks promising for the
computationally simple $\Delta$self-consistent field method.


%%%%%%%%%%%%%%%%%%%%%%%%%%%%%%%%%%%%%%%%%%%%%%%%%%%%%%%%%

\section{INTRODUCTION}

The development of powerful computers has allowed available {\it{ab
initio}} methods to calculate electronic excitations within
considerably larger systems than ever before\cite{Pople}. However, the
need to calculate electronically excited states within systems of such
a size, that the above methods are just not practical, is ever
present.

The less computational expansive density-functional
theory\cite{Hohenberg} (DFT) as proved its value as a theoretical
method over the years. Typically this method address only ground-state
properties. However, over the years there has been a number of studies
extending DFT into the realm of electronic excitations. Time-dependent
DFT\cite{Runge}, GW\cite{Hedin}, perturbation-DFT\cite{Gorling2} and
embedded DFT\cite{Kluner} methods, all show an impressive and
promissing progress in this field.

In this paper, DFT within a $\Delta$self-consistent field (SCF)
framework is used to calculate the ground and two lowest
electronically excited states of NaCl and LiCl. The DFT-based
$\Delta$SCF method, which recently has received some
justification\cite{Gorling} is an extraordinary simple method to
calculate electronic excited states. Here, state-of-the-art
multi-configuration self-consistent field (MCSCF)
method\cite{Shepard} is used to confirm the accuracy of the above
method by calculating the same electronic excitations.  The calculated
PES's obtained by both methods are compared both with other
theoretical results\cite{Zeiri} and experimental\cite{Silver}
ones. The overall agreement between the MCSCF method and the DFT-based
$\Delta$SCF method is promising for use of the $\Delta$SCF method on
systems of such size that are beyond the present realm of the MCSCF
method.\cite{Hellman}

As an extantion, the quantum dynamics of a simulated photodissociation
process is resolved using timedependent wavepacket propagation, on the
obtained PES's. The wavepacket is simultaneously propagated on the
coupled diabatic PES's and the distribution of the amplitudes are
monitored at each timestep so that the dissociation fraction can be
determined and also the intermediate dynamics of the photoreaction.

The organization of the paper is as follows. In section \ref{secII}
the model and the computational methods used in this paper are
described.  Results from the calculation are presented in section
\ref{secIII}. Conclusions are given in section \ref{secIV}. Finally,
the appendix state the results of a simulated photodissociation of the
two alkali-halide systems

%%%%%%%%%%%%%%%%%%%%%%%%%%%%%%%%%%%%%%%%%%%%%%%%%%%%%%%%%%%%%%%%%%%%
\section{THEORY AND COMPUTATIONAL METHOD} \label{secII}

The main advantage with an ab initio calculation, such as the MCSCF
method, is the predictive power in estimating the fundamental forces
that act on the involved nuclei, both in the ground state and in an
excited state.  Unfortunately, the highly accurate ab initio method is
hard to apply to larger systems, due to the high computational
cost. On the other hand, first-principle DFT has proven to be an
essential tool in describing large systems, whereas it has been
restricted to ground-state properties so far. However, recently
ordinary DFT has been extended to include electronically excited
states in a $\Delta$SCF-fashion,\cite{Gorling} with a working
accuracy.\cite{Hellman}

The optimized geometry was obtained by a restricted Hartree-Fock (RHF)
calculation using GAMESS.\cite{GAMESS} Here the basis set uses
Slater-type orbitals (STO) together with 6 Gaussians, hence STO-6G.
Both dimers have a $C_{2v}$ point group symmetry, in which the
Hamiltonian matrix is constructed in a basis that transforms according
to the $A_1(\Sigma^+,\Delta),B_1(\Pi)$, and $A_2(\Sigma^-,\Delta)$
irreducible representations. Here only the $\Sigma^+$ symmetry is
considered since one of the aims is to simulate a photodissociation
event. In addition, the use of symmetry reduces the complexity of the
calculation substantially. The wavefunction obtained from the RHF
calculation is used as an input to the MCSCF calculation, where 10
electrons are distributed in 14 active orbitals. This generate
approximately one million sets of configuration determinants for the
$A_1$ symmetry group.  The MCSCF determines the PES's for the ground
$^1\Sigma^+$ and the excited $^3\Sigma^+,^1\Sigma^+$ states with high
accuracy.

In analogy with the $\Delta$SCF\cite{Slater} method to calculate
electronically excited states within the Hartree-Fock (HF) approximation,
the DFT-based $\Delta$SCF method uses different electronic
configuration in the Kohn-Sham (KS) model system to represent these
electronically excited states. Such an application of the KS formalism
has for long been without any formal justification, but in a recent
article by A. G\"{o}rling\cite{Gorling} this method retrieves some
justification and indeed can be viewed as an approximative method to
calculate electronic excitations within the considered system.

The methodology of the used DFT-based $\Delta$SCF method is presented
in length elsewhere\cite{Hellman} and here only a short description is
given.  The basic ingredients are three concepts: (i) interpretation of
the KS-orbitals in a molecular orbital (MO) scheme, (ii)
discretization of these orbitals and their energy levels, using
supercell calculations with periodic boundary conditions, and (iii)
introduction of electron-hole (e-h) pairs in the system, which is
equivalent to an internal charge transfer in the supercell.  First an
ordinary DFT calculation is performed to obtain the ground state PES
of the system and the KS orbitals with discrete energy levels. Then
the relevant KS orbitals for the desired internal charge transfer
process are identified as the ones that should be occupied in the
ground state but unoccupied in the excited electronic configuration.
Next a hole is introduced in one of these identified occupied
KS orbitals together with an extra electron onto another one that is
introduced into the excited configuration. In this way a
KS determinant for the desired excitation is constructed. Finally,
this KS determinant is optimized in a self-consistent-field
calculation, and its energy is evaluated as in a normal DFT
calculation.  The total energy difference between the excited- and
ground-state electronic configurations is identified as the excitation
energy.

This method is straightforward, when calculating the covalent triplet
$^3\Sigma$ state of the NaCl and LiCl dimers, since it can be
constructed using only one KS determinant. However, in the
photodissociation process the transitions from the ionic ground-state
$^1\Sigma$ to the excited state $^3\Sigma$ is forbidden. Therefore we
apply the so called ``sum-method''\cite{Ziegler} to calculate the
singlet $^1\Sigma$ state of the NaCl and LiCl dimers. Simply, it means
that a weighted sum of determinants is constructed, including both the
$^3\Sigma$ and $^1\Sigma$ state, and its energy is found by means of
ordinary minimization procedure. After that, the energy for the
singlet state can be extracted since the sole energy for the triplet
state can be found by similar means.

The first principle calculations presented in this paper are performed
by means of the plane wave pseudopotential code {\tt
  DACAPO}~\cite{DACAPOCODE}. The generalized-gradient (GGA)
approximation~\cite{PerdewI,PerdewII,PerdewIII} is used for the
exchange-correlation energy-density functional. The wave functions are
expanded in a plane-wave basis set, and the electron-ion interactions
are described by ultrasoft pseudopotentials.  The electronic density
between iterations is updated by means of a Pulay-mixing algorithm.
The occupation numbers are updated using a developed technique based
on minimization of the free energy functional.  All calculations are
performed allowing for spin polarization.  The dimer cases LiCl and
NaCl, are calculated using a supercell of a volume of
20$\times$20$\times$20 \AA$^3$. The reason for using such a big
supercell is to minimize any artificial effect from the periodicity.

%%%%%%%%%%%%%%%%%%%%%%%%%%%%%%%%%%%%%%%%%%%%%%%%%%%%%%%%%%%%%%%%%%%%

\section{Result and Discussion}\label{secIII}

Here the results for total energies and aspects of the electronic
structure, such as the charge density, are presented.
%################################################################
\begin{figure}[h]
\resizebox{7cm}{!}{\includegraphics{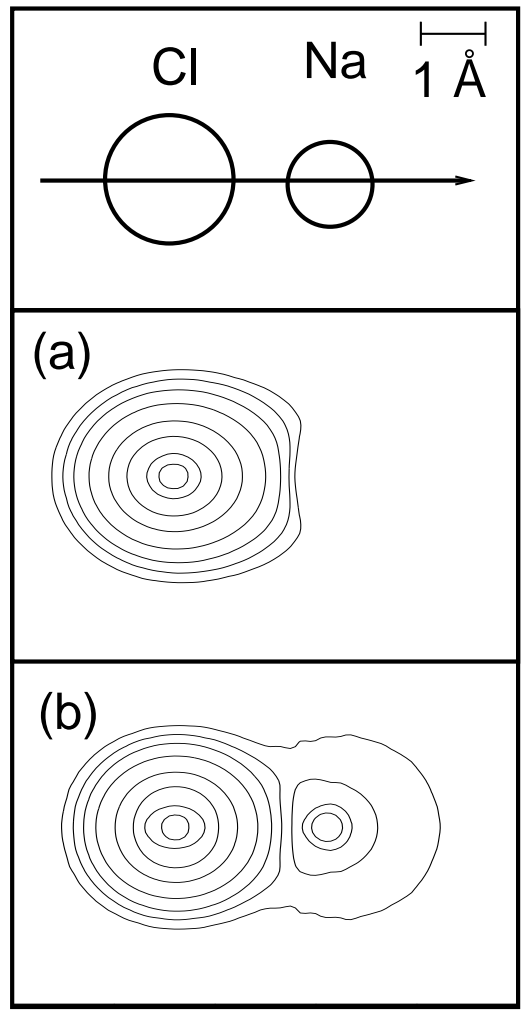}}
\vspace{0.2cm}
\caption{The different valence electron density profiles of the NaCl molecule through a cut along the molecular axis. (a) The ground state with its ionic character, (b) The covalent $^3\Sigma$ state between Na and Cl.} 
\protect{\label{fig:1}}  
\end{figure}
%################################################################ 
The electronic ground states for the NaCl and LiCl dimers at their
equilibrium distances are ionic, with the Na(Li) and Cl atoms being
positively and negatively charged, respectively. However, at infinite
separation the ground state is a neutral ``covalent'' configuration,
where the ionic configuration lies about $1.4(1.7)$ eV above the
covalent PES. This value comes from the difference in the ionization
energy for the Na(Li) atom and the affinity energy of the Cl atom. As
the Na(Li) and Cl atom move toward each other, there occurs at some
intermediate separation ($\sim$ 10 \AA)\cite{Zeiri} approximative the
PES's for the covalent and ionic states have comparable energies,
which makes an internal charge transfer of the unpaired 3(2)s electron
on the Na(Li) to the electronegative Cl atom and the electrostatic
force affect the ionic fragment more strongly than in the original
weak covalent bond. The ionic character of the ground state can be
seen in Fig.~\ref{fig:1}a where the cut through the dimer axis clearly
shows how the charge is focused around the chloride atom.

Analyzing the DOS for the system shows that the electron transfer
moves the $3s$-electron from the sodium atom to the empty $3p^6$
orbital on the chloride atom, as expected. So in order to obtain the
excited triplet $^3\Sigma$ covalent state, the hole is introduce in
the now filled $3p^6$ orbital and the electron in the parallel
spin-channel of the $3s$ orbital of the Na atom. Figure~\ref{fig:1}b
shows the covalent character of the excited $^3\Sigma$ state with a
charge density located around the sodium atom, representing the $3s$
electron, and more importantly, a concentration of charge between the
atoms. Since we are interested in the excited singlet $^1\Sigma$
state, we construct a combination of the triplet state and the singlet
state by calculating the energy for the KS-determinant constructed by
instead placing the electron in the anti-parallel spin-channel of the
$3s$ orbital of the Na atom. The energy separation of the
triplet-singlet state is then extracted from this calculation.

\begin{figure}[h]
\resizebox{8cm}{!}{\includegraphics{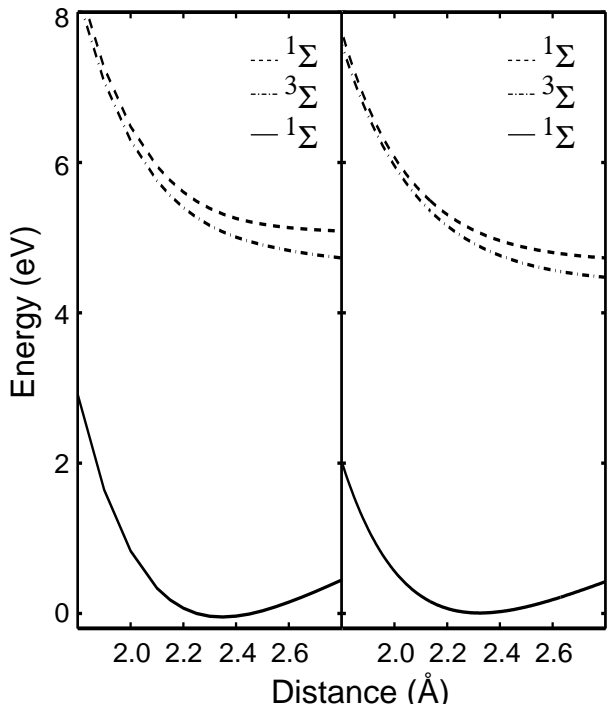}}
\vspace{0.2cm}
\caption{The calculated potential energy curves for the NaCl system with the MCSCF (left) and $\Delta$SCF methods (right). The ionic $^1\Sigma$ ground state is displayed in a solid line while the excited $^3\Sigma,^1\Sigma$ covalent states uses a dashed line.} 
\protect{\label{fig:2}}  
\end{figure}

In Fig.~\ref{fig:2} the PES's for the ionic $^1\Sigma$ ground state
and both covalent $^3\Sigma,^1\Sigma$ excited states of the NaCl dimer
are shown, calculated with the MCSCF method and DFT-based $\Delta$SCF
method. 
%################################################################
%################################################################
In Table 1 the bondlength, and vertical excitation energies calculated
with the present methods are compared with values from other
calculations\cite{Zeiri} and experiment,\cite{Silver} for both the
NaCl and LiCl dimers.  The overall agreement between the point from
the MCSCF and $\Delta$SCF methods is promising. For instance, the
discrepancy in the (i) vertical excitation for a $^1\Sigma\rightarrow
^1\Sigma$ transition for both the NaCl and LiCl dimers is around 0.3
eV, (ii) singlet-triplet splitting is around 0.05 eV, and (iii)
bondlength is around 0.03 \AA\ .

\begin{table}
\caption{The calculated bondlength for the ionic $^1\Sigma$ state and vertical excitation energies to both $^3\Sigma,^1\Sigma$ states, for the NaCl and LiCl dimers. Here results from (a) the MCSCF method, (b) the $\Delta$SCF method, are compared with those for (c) the valence-bond (VB) method, and experiment.  The unit is eV.} 

\begin{center}
\begin{tabular*}{17cm}{l l l l l }
NaCl                   & MCSCF  & $\Delta$SCF & VB  & Exp.   \\
\hline
Bondlength                   & 2.35 & 2.32 & 2.34 & 2.36   \\ 
$^1\Sigma \rightarrow ^3\Sigma$& 5.27 & 4.92 & -    & -      \\
$^1\Sigma \rightarrow ^1\Sigma$& 5.49 & 5.17 & 5.74 & 5.26   \\
\hline\hline
LiCl                   &  &  &  &    \\
\hline
Bondlength                   & 2.02 & 2.06 & 2.07 & 2.02   \\ 
$^1\Sigma \rightarrow ^3\Sigma$& 5.96 & 5.64 & -    & -      \\
$^1\Sigma \rightarrow ^1\Sigma$& 6.08 & 5.75 & 7.28 & -      \\
\end{tabular*}
\end{center}
\label{table2}
\end{table}

%%%%%%%%%%%%%%%%%%%%%%%%%%%%%%%%%%%%%%%%%%%%%%%%%%%%%%%%%%%%%%%%%%%%%%
\section{CONCLUSIONS}\label{secIV}

The ground and two lowest excited states of NaCl and LiCl dimers are
calculated.  In the paper there are two main theoretical methods used;
(i) the ab initio MCSCF method with high accuracy and (ii) a
first-principle DFT-based $\Delta$SCF method with a working accuracy
to calculate both the ionic $^1\Sigma$ ground state and the two
$^3\Sigma,^1\Sigma$ covalent states of the systems.

The overall agreement between results of the MCSCF and DFT-based
$\Delta$SCF methods is very promising. For instance, the discrepancy in
vertical excitation for a $^1\Sigma\rightarrow ^1\Sigma$ transition
for both the NaCl and LiCl dimer is around 5$\%$ and in bondlength
around 2$\%$ compared to each other.

%%%%%%%%%%%%%%%%%%%%%%%%%%%%%%%%%%%%%%%%%%%%%%%%%%%%%%%%%%%%%%%%%%%%%%

\section{ACKNOWLEDGMENTS}
The work is supported by the Swedish Scientific 
Council and the Swedish Foundation for Strategic Research
(SSF) via Materials Consortia No.~9 and ATOMICS, which is gratefully
acknowledged. We thank B.~I.~Lundqvist for comments on the manuscript.

%%%%%%%%%%%%%%%%%%%%%%%%%%%%%%%%%%%%%%%%%%%%%%%%%%%%%%%%%%%%
\section{APPENDIX}

Knowing the PES's for the system enables the use of wavepacket
propagation to resolve the quantum dynamics of the nuclei.  
However, the quantum dynamics of the photodissociation process on its multitude
of diabatic potentials is a complex problem.  As the system makes a
sudden transition from its ground state to its excited state there
will be a new set of forces that work on the dimer. The potential
energy will be released into the intermolecular coordinate which might
result in a dissociation of its fragments.  The timeevolution of the
system is determined by the Schr\"{o}dinger equation
\begin{equation}
i\hbar\frac{\partial \Psi}{\partial t}={\hat{H}}\Psi,
\end{equation}
where the Hamiltonian matrix for the NaCl dimer looks like
\begin{equation}
{\hat{H}}=\left[ \begin{array}{ccc} {\hat{H}}_{cov}& &  {\hat{V}}_{12} \\
                                  {\hat{V}}_{21}& & {\hat{H}}_{ion} 
\end{array}.
\right]
\end{equation}
Here the diagonal elements $\hat{H}_{cov}$ and $\hat{H}_{ion}$
represent well-defined electronic configurations of the NaCl and LiCl
dimers, i.e.~the ionic ground state and the covalent excited state.
The off-diagonal element ${\hat{V}}_{21}$ is the non-adiabatic
coupling between the diabatic states, which enables the wavepacket to
bifurcate among the states.  It has been found\cite{Grice} that the
strength of the non-adiabatic coupling term ${\hat{V}}_{21}$ depends
exponentially on the position of the curve-crossing point $R_{cross}$
as
\begin{equation}
{\hat{V}}_{21}=V_{12}\exp[\gamma R_{cross}],
\end{equation}
where for the halide Cl atom the parameters are $V_{12}=20$ eV,
$\gamma=1.1638$~\AA$^{-1}$. 

The solution to the Sch\"{o}dinger equation is calculated with a
timedependent wavepacket method that is based on discrete variables and a
finite basis representation (DVR-FBR).\cite{Kosloff}  The DVR has the
advantage that the amplitude of the wavepacket is well defined, and it
leads itself to the simultaneous propagation of the wavepacket on
different diabatic potentials. The propagation of the wavepacket is
done with the standard split operator technique\cite{Feit} where the
potential operator together with the effect of the non-adiabatic
coupling are evaluated in the DVR, while the kinetic operator is
calculated in the FBR.

%################################################################
\begin{figure}[ht]
\resizebox{7cm}{!}{\includegraphics{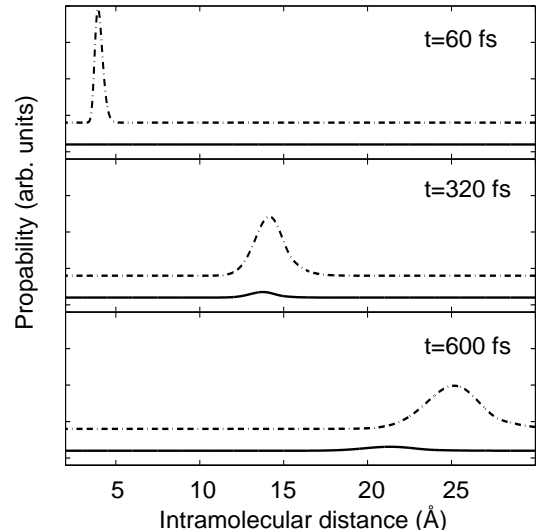}}
\vspace{0.2cm}
\caption{The timeevulotion of $|\langle\Psi(t)|\Psi(t)\rangle|^2$, representing the probability distribution for the NaCl dimer, at three different times. Here the dashed-dotted curve shows the probability distribution on the covalent state, whereas the solid curve is the probability distribution of the ionic state.}
\protect{\label{fig:3}}
\end{figure}
%################################################################
%################################################################
\begin{figure}[h]
\resizebox{7cm}{!}{\includegraphics{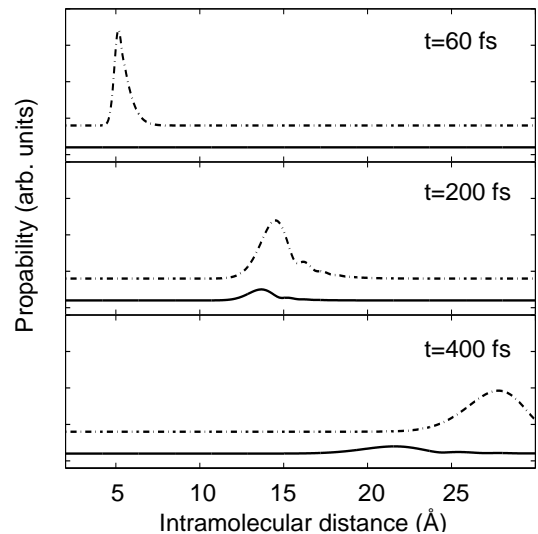}}
\vspace{0.2cm}
\caption{The timeevulotion of $|\langle\Psi(t)|\Psi(t)\rangle|^2$, representing the probability distribution for the LiCl dimer, at three different times. Here the dashed-dotted curve shows the probability distribution on the covalent state, whereas the solid curve is the probability distribution of the ionic state.}
\protect{\label{fig:4}}
\end{figure}
%################################################################

As a consequence of the overall good agreement between the PES's
obtained by the MCSCF and DFT-based $\Delta$SCF methods, the results
from the quantum dynamics turns out to be the same within the
simulations accuracy. Here the quantum dynamics performed by the NaCl
dimer during photodissociation is found to be highly non-adiabatic,
with a negligible population of the quasi-bound state, whereas the
LiCl dimer has a small population of the quasi-bound state.  It is
concluded that the population of the quasi-bound state strongly
depends on the difference between the ionization potential and the
affinity for the two atoms involved in the dimer. Hence, if one moves
up the rows of the alkali-atoms or down the rows of the halogen-atoms
the population of the quasi-bound state is expected to increase
substantially. The observation of oscillations in pump-probe
experiments on NaI dimers\cite{Rose} should be consistent with this
conclusion.

% Choose bibliographystyle. prsty_title is edited
% to include the title of articles (both in journals
% and inbooks) 

%\linebreak

%%%%%%%%%%%%%%%%%%%%%%%%%%%%%%%%%%%%%%%%%%%%%%%%%%%%%%%%%%%%

%%%%%%%%%%%%%%%%%%%%%%%%%%%%%%%%%%%%%%%%%%%%%%%%%%%%%%%%%%%%

%%%%%%%%%%%%%%%%%%%%%%%%%%%%%%%%%%%%%%%%%%%%%%%%%%%%%%%%%%%%

%%%%%%%%%%%%%%%%%%%%%%%%%%%%%%%%%%%%%%%%%%%%%%%%%%%%%%%%%%%%
\end{document}